# The General Quantum Limit for and the Optimization of the Minimum Measurable Frequency Shift in a Laser

Selim M. Shahriar[1,2], Ruoxi Zhu[1], Jinyang Li[2], Zifan Zhou[1], Dustin Greenwood[2], Yael Sternfeld[3], Michal Hazan[4], Yehonatan Zolti[4], Jacob Scheuer[4], and David Smith[5]

[1]*Department of Electrical and Computer Engineering, Northwestern University, Evanston, IL 60208, USA*

[2]*Department of Physics and Astronomy, Northwestern University, Evanston, IL 60208, USA*

[3]*Department of Physics, University of California, Berkeley, CA 94720, USA*

[4]*School of Electrical Engineering, Faculty of Engineering, Ramat-Aviv, Tel Aviv 69978, Israel*

[5]*NASA Marshall Space Flight Center, Space Systems Department, ES23, Huntsville, Alabama 35812, USA*

Abstract

We show that, contrary to conventional understanding, the minimum measurable frequency shift (MMFS) for a single-mode ideal laser is determined by a combination of phase diffusion caused by SE and the shot noise caused by the VM. For all practical sensors, the MMFS is found to be given by the geometric mean of the measurement bandwidth and the Schwalow-Townes Linewidth, multiplied by a factor which can be much larger than unity under certain conditions. We determine the optimal value for the MMFS for three different sensing modalities, an unbalanced Mach-Zehnder Interferometer, a passive Fabry-Perot cavity (FPC), and heterodyning with a reference laser, and identify the conditions needed for reaching the optimal value.

## 1. Introduction

Many applications in the arena of precision metrology are based on measurement of the shift in the frequency of a laser. The sensitivity of such a measurement is quantified by the minimum measurable frequency shift (MMFS) for a given measurement time. If a resonator, such as a Fabry-Perot cavity (FPC), is employed to determine the effect of some measurand, the MMFS for the source laser also places a lower bound on the minimum measurable change in its resonance frequency. One of the most important of such applications is the measurement of rotation, via the Sagnac effect, using a pair of counter-propagating ring lasers [1,2,3,4,5,6,7]. Such gyroscopes are routinely used for inertial navigation. For a large enough area, such a gyroscope may be able to reach a degree of accuracy high enough for precision test of a prediction of General Relativity, namely the gravitational frame dragging effect, by measuring the Lens-Thirring rotation [8,9,10]. Another important application is the detection of ultra-light dark matter, which is predicted to produce a time-varying change in the frequency of a laser or the resonance frequency of an FPC [11,12].

There are fundamentally two types of techniques for measuring the frequency shift in a laser. The first one is heterodyne detection, where one measures the spectrum of the beat frequency between the test laser and a reference laser. The second one is time-delayed homodyning, where the output of the test laser is made to interfere with a time-delayed part of its output. There are multiple ways to implement this approach. One method is to use an unbalanced Mach-Zehnder interferometer (UMZI) or an unbalanced Michelson interferometer (UMI). In either case, variation of the frequency of the test laser produces a fringe with its width determined by the difference between the two arm lengths. Measuring a change in the peak of this fringe, for example, allows one to determine a shift in the laser frequency. Another method is to use an FPC, in which the output signal results from interference among multiple bounces, each time-delayed with the initial beam. Measuring a change in the peak of the FPC resonance allows one to determine a shift in the laser frequency.

For the heterodyning technique, it has been claimed [13,14,15,16] that the MMFS is given by a characteristic frequency, $\Gamma_0 \equiv \sqrt{\gamma_m \gamma_{STL}}$ , where $\gamma_m$ is the measurement bandwidth (full-width-half-max: FWHM), and $\gamma_{STL}$ is the Schawlow-Townes linewidth (STL) (FWHM). In ref. 14, the authors show that this limit is due to the phase-diffusion caused by SE (SE) into the lasing mode. For the time-delayed homodyning technique, there is no published analysis, to the best of our knowledge, for the MMFS employing an UMZI or an UMI. However, for an FPC, it has generally been accepted that the MMFS is given by the ratio of the spectral width of the resonance and the signal-to-noise ratio (SNR), dictated by the shot-noise [17]. In ref. 14, the authors derive an expression for the MMFS for an FPC using the fundamental uncertainty relations between the number of photons detected and the phase. However, as they have pointed out, the result is the same as the one mentioned above: the ratio of the spectral width of the resonance and the shot-noise limited SNR.

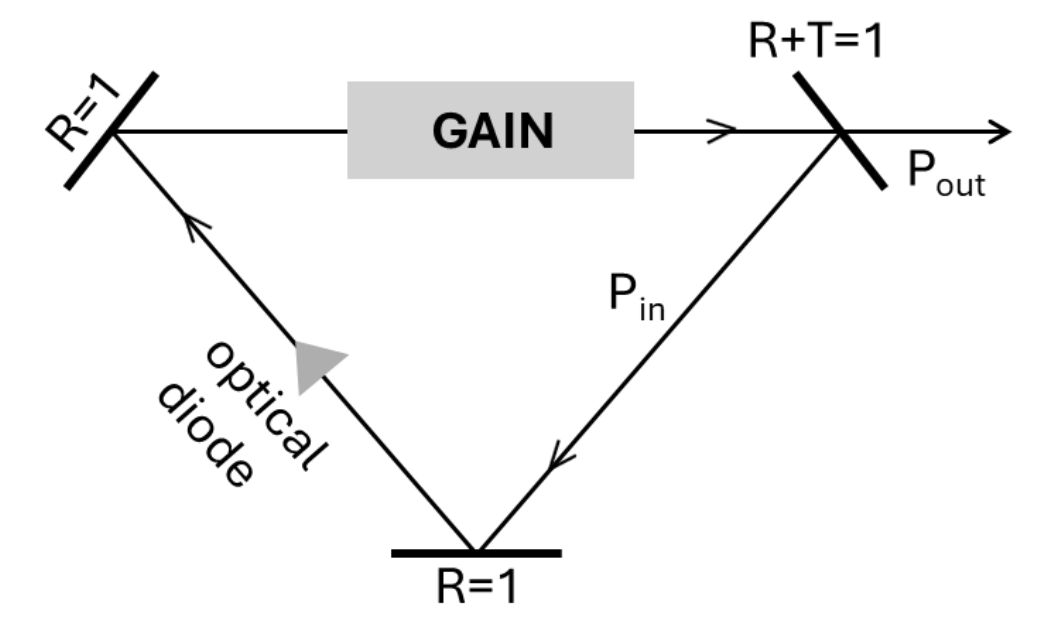

Figure 1: Schematic illustration of a single mode ring laser, enclosed by two perfect reflectors and one output coupler. The optical diode ensures uni-directional lasing.

Here, we show that these limits are incorrect. Specifically, we show that for all sensors the MMFS is due to a combination of the phase diffusion due to spontaneous emission (SE) into the lasing mode, and the shot-noise caused by the vacuum mode (VM) that accompanies a coherent state. Specifically, the MMFS is found to be $\xi\Gamma_o$ , with $\xi$ being much greater than unity in some cases. We also show that for optimal choice of parameters the value of $\xi$ can be minimized to the value of $\sim\sqrt{1+(\gamma_{STL}/\gamma_{LC})^2}$ , where $\gamma_{LC}$ is the laser cavity linewidth, and identify the conditions needed for reaching this value for all sensors.

The rest of the paper is organized as follows. In Section 2, we develop the general model for the MMFS of a single-mode ideal laser. In Section 3, we consider three different modalities for measuring frequency shifts, and identify conditions for minimizing the MMFS for each modality. We summarize the findings in Section 4. In the Appendix, we show that the general model for the MMFS is consistent with the Cramer-Rao bound.

## 2. General model for the MMFS of a laser

In deriving the MMFS, we consider for specificity a continuous wave, single mode laser, operating in a steady state. For concreteness of discussion, we assume that the lasing occurs unidirectionally in a ring cavity, as illustrated in Figure 1. We also assume that its spectral width is limited by quantum noise only. Under these assumptions, the MMFS is determined by two independent effects: the phase diffusion caused by SE into the lasing mode, and the shot-noise caused by the VM accompanying a coherent state. In what follows, we first consider the effects of these two processes separately, and then determine the combined effect on the MMFS.

### 2.1 Effect of Spontaneous Emission on the MMFS of a Laser

Consider the effect of spontaneous emission (SE) into the lasing mode of the cavity. The rate of this emission can be found as follows. The number of photons exiting the laser cavity during its decay time can be expressed as $\langle n\rangle_{LC} \equiv P_{out}\tau_{LC}/\hbar\omega_o$, where $P_{out}$ is the output power, $\tau_{LC}$ is the lifetime of the laser cavity, and $\omega_o$ is the lasing frequency. Therefore, $\langle n\rangle_{LC}$ represents the number of photons in the laser cavity in steady state. It also follows that the rate of stimulated emission necessary to maintain this photon number can be expressed as $\Gamma_{STE} = \langle n\rangle_{LC}/\tau_{LC}$. Consider next the rate of SE into the lasing mode, $\Gamma_{SPE,LM}$. Since this emission rate is proportional to the single photon in the VM, while $\Gamma_{STE}$ is proportional to $\langle n\rangle_{LC}$, it follows that $\Gamma_{SPE,LM}/\Gamma_{STE} = 1/\langle n\rangle_{LC}$. Denoting the laser cavity line-width (half-width-half-max: HWHM) as $\gamma_{LC}$, we see that $\Gamma_{SPE,LM} = 1/\tau_{LC} = \gamma_{LC}$.

We express the positive frequency part of the laser field outside the cavity as $E^+(t) = E_o \exp\left[-i\omega_o t + i\theta(t)\right]$. The SE into the lasing mode cause phase diffusion for the laser, so that $\theta(t)$ is a random variable. Under stationary conditions, the phase diffusion can be characterized by the difference between this phase at two different times: $\phi_\tau \equiv \theta(t+\tau)-\theta(t)$. Using a random walk model for this process, it can be shown that [18] this stationary phase difference obeys the following probability distribution:

$$P(\phi_\tau) = \exp\left[-(\phi_\tau/\sigma)^2\right]/\sqrt{\pi\sigma^2};\ \sigma^2 = 2\tau/\tau_{STL} \tag{1}$$

where $\tau_{STL} = 1/\gamma_{STL} = 2\langle n\rangle_{LC}/\gamma_{LC}$. When integrated over this symmetric distribution, it is easy to show that $\langle \exp(i\phi_\tau)\rangle = \exp(-\tau/2\tau_{STL})$. The spectrum of the laser is given by the Fourier transform of the two-time correlation function of the electric field. As such, it follows that the spectral width (i.e., the Schawlow-Townes Linewidth) of the laser is given by $\gamma_{STL} = 1/\tau_{STL} = \gamma_{LC}/2\langle n\rangle_{LC}$.

Consider a frequency measurement carried out over a time duration of $\tau_S = 1/\gamma_S$, where $\gamma_S$ is the spectral width of the measurement system. The standard deviation of the random phase can be determined immediately from Eqn. (1) to be $\Delta\phi = \sigma/\sqrt{2} = \sqrt{\tau_S/\tau_{STL}}$. Therefore, the frequency uncertainty is given by $\Delta\omega = \Delta\phi/\tau_S = \left(\sqrt{1/\tau_S\tau_{STL}}\right)$. If the measurement is carried out $M \equiv \tau_M/\tau_S = \gamma_S/\gamma_M$ times, the frequency uncertainty is reduced by a factor of $\sqrt{M}$ [19] The MMFS, due to the effect of the SE only, is thus given by:

$$\Delta\omega_{MIN,SE} = \sqrt{1/(\tau_S\tau_{STL})}\Big/\sqrt{\tau_M/\tau_S} = \sqrt{1/(\tau_M\tau_{STL})} = \sqrt{\gamma_M\gamma_{STL}} = \Gamma_0. \tag{2}$$

As we show later, this result is strictly valid when $\tau_S \ll \tau_{STL}$, which is typically the case for any system of practical interest. Under this constraint, this result is universal, independent of the technique used for measuring frequency shift, or the condition used for doing so. Note that this constraint does not limit the maximum value of the overall measurement time, $\tau_M$. When this constraint does not hold, the MMFS due to SE may be somewhat larger than this value. As such, Eqn. (2) represents the optimal value of the MMFS due to SE.

### 2.2 Effect of the Vacuum Mode on the MMFS of a Laser

For a coherent state, $|\alpha\rangle$, it is easy to show that the mean value of the number of photons is given by $\langle \hat{N} \rangle = |\alpha|^2 \equiv \langle n \rangle$, while $\langle \hat{N}^2 \rangle = |\alpha|^2 + |\alpha|^4$, so that the uncertainty in the number of photons is $\Delta N = |\alpha| = \sqrt{\langle n \rangle}$. This uncertainty, which is generally called the shot-noise, is directly attributable to the fact that the coherent state is a displaced vacuum mode (VM). In other words, the laser field in a coherent state can be viewed as the sum of a deterministic mean field and a field with a random phase distributed evenly over $\{0, 2\pi\}$ which represents the VM.

To see how the presence of this mode affects the MMFS, consider a frequency measurement carried out over a time duration of $\tau_S = 1/\gamma_S$, where $\gamma_S$ is the spectral width of the measurement system. If we ignore for now the effect of the phase diffusion, the number of photons detected during this time is given by:

$$n = P_{out} \tau_S / \hbar\omega_o = \langle n \rangle_{LC} \tau_S / \tau_{LC} = \langle n \rangle_{LC} \gamma_{LC} / \gamma_S \tag{3}$$

The corresponding uncertainty in the number of photons is given by $\Delta n = \sqrt{\langle n \rangle_{LC} \gamma_{LC} / \gamma_S}$.

The fundamental uncertainty relation between the phase and the photon number can be expressed as:

$$\Delta\phi \cdot \Delta n \geq 1/2 \tag{4}$$

The equality limit for this expression holds only in the limit where the phase diffusion caused by SE is vanishingly small, so that the field detected for measuring the frequency shift is in a pure coherent state. In this limit, we get $\Delta\phi = 1/2\Delta n = \sqrt{\gamma_S / 4\langle n \rangle_{LC} \gamma_{LC}}$. To determine the frequency, the phase has to be measured twice. As such, the corresponding frequency uncertainty is given by $\Delta\omega = \sqrt{2}\Delta\phi/\tau_S$. (Note that, for the case of the phase diffusion caused by SE, this additional factor of $\sqrt{2}$ is already taken into account by equating the phase uncertainty to the standard deviation). If the measurement is carried out $M = \tau_M/\tau_S = \gamma_S/\gamma_M$ times, the frequency uncertainty is reduced by a factor of $\sqrt{M}$ [20]. The value of the MMFS due to effect of the VM only is thus found to be:

$$\Delta\omega_{MIN,VM} = \frac{\sqrt{2}\Delta\phi}{\tau_S} \frac{1}{\sqrt{M}} = \gamma_S \sqrt{\frac{\gamma_S}{2\langle n \rangle_{LC} \gamma_{LC}}} \sqrt{\frac{\gamma_M}{\gamma_S}} = \gamma_S \sqrt{\frac{\gamma_M \gamma_{STL}}{2\gamma_{LC}^2}} = \varepsilon_o \Gamma_0; \quad \varepsilon_o = \gamma_S / \gamma_{LC} \tag{5}$$

One must make use of optimal measurement conditions to reach this limit for $\Delta\omega_{MIN,VM}$. For example, in the case of the UMZI, the measurement is to be carried out at the point where the slope of the fringe is maximum in order to reach this value for $\Delta\omega_{MIN,VM}$, as we show in Section 3.1, as well as in the Appendix.

### 2.3 Overall MMFS of a Laser

Consider the signal, denoted as $S$, that is monitored in order to measure the frequency shift of a test laser. The MMFS can be expressed as:

$$\Delta\omega_{MIN} = \left|\frac{\Delta S}{\partial S/\partial\omega}\right| \tag{6}$$

where $\Delta S$ is the standard deviation of the signal, which is given by the square-root of the variance of the signal, $\Delta S^2$, and $S$ is the mean value of the signal. This expression is valid as long as the slope of the signal, $\partial S/\partial\omega$, remains nearly constant over the range of $\omega$ corresponding to $\Delta\omega_{MIN}$. This condition generally hold for practical systems of interest. For simplicity, but without loss of generality, we assume that $S$ is dimensionless, and represents the number of photons detected. In the limit where the fluctuations in the signal caused by the SE and the VM can be approximated to be independent of each other, we can write that:

$$\Delta\omega_{MIN,SE} = \left|\frac{\Delta S_{SE}}{\partial S/\partial\omega}\right|;\ \Delta\omega_{MIN,VM} = \left|\frac{\Delta S_{VM}}{\partial S/\partial\omega}\right|;\ \Delta\omega_{MIN} = \left|\frac{\sqrt{\Delta S_{SE}^2 + \Delta S_{VM}^2}}{\partial S/\partial\omega}\right| = \sqrt{\Delta\omega_{MIN,SE}^2 + \Delta\omega_{MIN,VM}^2} \tag{7}$$

In Section 3.1, we show, for the case of the UMZI, that when the measurement is carried out at the point where the slope is maximum, the values of $\Delta\omega_{MIN,SE}$ and $\Delta\omega_{MIN,VM}$ as expressed in Eqn. (7) agree exactly with the corresponding values shown in Eqn. (2) and Eqn. (5), respectively.

The MMFS due to the VM can be expressed as:

$$\Delta\omega_{MIN,VM} = \left|\frac{\sqrt{S}}{\partial S/\partial\omega}\right| \tag{8}$$

since the shot-noise is given by the square root of the mean number of photons, even in the presence of phase diffusion caused by SE. Again, this expression is valid as long as the slope of the signal, $\partial S/\partial\omega$, remains nearly constant over the range of $\omega$ corresponding to $\Delta\omega_{MIN,VM}$.In general, the value of the MMFS due to the VM would depend on the operating condition for a given measurement scheme. For example, in the case of the UMZI, this MMFS is optical when the measurement is carried out at the point where the slope of the signal is maximum, as shown in Section 3.1.

Consider next the MMFS due to SE, for a single measurement over the time duration of $\tau_S$. This can be expressed as:

$$\Delta\omega_{MIN,SE} = \left|\frac{\Delta S_{SE}}{\partial S/\partial\omega}\right| = \left|\frac{\Delta\phi\bullet(\partial S/\partial\phi)}{(\partial S/\partial\phi)(\partial\phi/\partial\omega)}\right| = \frac{\Delta\phi}{\tau_S} = \frac{\sqrt{\tau_S/\tau_{STL}}}{\tau_S} = \frac{1}{\sqrt{\tau_S\tau_{STL}}} \tag{9}$$

If the measurement is carried out $M \equiv \tau_M / \tau_S = \gamma_S / \gamma_M$ times, the frequency uncertainty is reduced by a factor of $\sqrt{M}$ , so that we get:

$$\Delta\omega_{MIN,SE} = \frac{1}{\sqrt{\tau_S \tau_{STL}}} \frac{1}{\sqrt{\tau_M / \tau_S}} = \frac{1}{\sqrt{\tau_M \tau_{STL}}} = \sqrt{\gamma_M \gamma_{STL}} = \Gamma_0 \tag{10}$$

This expression is the same as the one shown earlier in Eqn. (2). However, we have now derived this result from the fundamental definition of the MMFS, as expressed in Eqn. (6). From this derivation, it is clear that, unlike the MMFS for the VM, the MMFS for SE is independent of the operating point, as long as the slope of the signal remains essentially constant over the range of frequencies corresponding to the MMFS.

We can thus write the overall MMFS as:

$$\Delta\omega_{MIN} = \sqrt{\Delta\omega_{MIN,SE}^2 + \Delta\omega_{MIN,VM}^2} = \sqrt{\Gamma_0^2 + \left|\frac{\sqrt{S}}{\partial S/\partial\omega}\right|^2} = \Gamma_0\sqrt{1 + \left|\frac{\sqrt{S}}{\Gamma_0 \partial S/\partial\omega}\right|^2} \tag{11}$$

In the Appendix, we present an alternative formulation of the derivation of the overall MMFS of a laser, for the special case of the UMZI. We show that the MMFS in that case, derived using two different approaches, namely error propagation and application of the Cramer-Rao bound (CRB), yield matching results when the mean number of photons is large. We also show that this model agrees with the approximations made here in the proper limits. It is difficult to carry out the same type of modeling for other cases, such as heterodyning and the FPC. However, since the approximate model presented here agrees with the more rigorous model in the proper limits, one would be justified in using the approximate model for these cases as well.

## 3. Optimization of the MMFS for specific sensing modalities

As noted above, the effect of the phase diffusion due to SE is essentially universal, and independent of the modality used for measuring the frequency shift in any laser in the experimentally relevant limit of $\tau_S << \tau_{STL}$ . However, that is not the case for the effect of the VM. In what follows, we consider three different modalities for measuring the frequency shift, and show how to minimize the effect of the VM in each case.

### 3.1 Minimizing the MMFS for a Laser Employing a UMZI or a UMI

One approach for detecting the frequency shift of a laser is to employ an unbalanced Mach-Zehnder interferometer (UMZI), illustrated schematically in Figure 2. In the UMZI, the two arms differ in length by $L_o$. Due to this imbalance, the signal observed in either output port varies sinusoidally as a function of the laser frequency. Any shift in this signal can be used to determine the corresponding change in the laser frequency. The same process can be realized using an unbalanced Michaelson interferometer (UMI), with a finite difference in the lengths of the two arms. The analyses for the MMFS of the laser in these cases are essentially identical. As such, we will discuss here the case of the UMZI only.

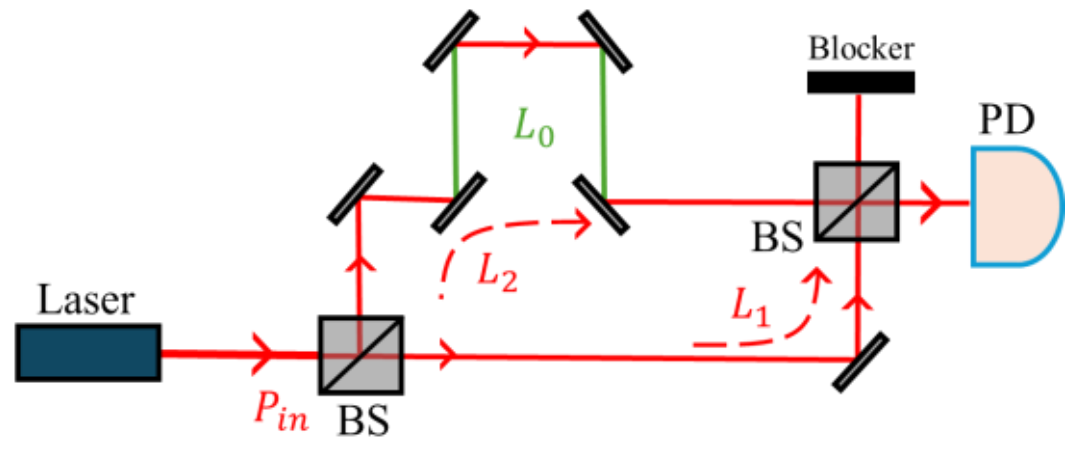

Figure 2: Schematic illustration of an unbalanced Mach-Zehnder Interferometer. See text for details.

We denote the time delay as $\tau_D = L_o / c$, and the number of photons emerging from the laser in a measurement time of $\tau_S$ to be $S_o = P_{out}\tau_S / \hbar\omega_o$. The delay time here plays the role of $\tau_S = 1/\gamma_S$, defined in Section 2.2, so that the spectral width of the measurement system is $\gamma_S = 1/\tau_D$. In the limit where the effect of the phase diffusion caused by SE is negligible, the signal, in terms of the number of photons arriving at the detector, can be expressed as $\tilde{S} = [1+\cos(\tau_D\omega + \phi_o)]S_0/2$, where $\phi_o$ represents an offset phase difference between the two arms. When the effect of the phase diffusion due to SE is taken into account, the signal is expressed as $\tilde{S} = [1+\cos(\tau_D\omega + \phi_o + \phi_{\tau_D})]S_0/2$, where $\phi_{\tau_D}$ is a random variable that obeys the probability distribution shown in Eqn. (1). When the signal is averaged over this distribution, we get:

$$S \equiv \langle \tilde{S} \rangle = [1+\exp(-\tau_D/2\tau_{STL})\cos(\tau_D\omega + \phi_o)]S_o/2 \tag{12}$$

From Eqn. (9), with $\tau_S = \tau_D$, and assuming that $\tau_D << \tau_{STL}$, it follows that the MMFS due to the SE is given by:

$$\Delta\omega_{MIN,SE} = \frac{1}{\sqrt{\tau_D\tau_{STL}}} \tag{13}$$

If the measurement is carried out $M \equiv \tau_M/\tau_D$ times, the frequency uncertainty is reduced by a factor of $\sqrt{M}$, so that we get:

$$\Delta\omega_{MIN,SE} = \frac{1}{\sqrt{\tau_D\tau_{STL}}}\frac{1}{\sqrt{\tau_M/\tau_D}} = \frac{1}{\sqrt{\tau_M\tau_{STL}}} = \sqrt{\gamma_M\gamma_{STL}} = \Gamma_0 \tag{14}$$

The expression for the MMFS due to the VM follows from Eqn. (8). However, we can now choose the operating point to minimize its value. One option is to operate at the point where the fringe has the highest slope, corresponding to $\phi_0 = \pm\pi/2$. For this operating point, we get

$$\Delta\omega_{MIN,VM} = \left|\frac{\sqrt{S}}{\partial S/\partial\omega}\right|_{\phi_0=\pm\pi/2} = \frac{1}{\sqrt{2S_0}}\cdot\frac{1}{\tau_{STL}}\cdot\frac{e^X}{X}; \quad X \equiv \frac{\tau_D}{2\tau_{STL}} \tag{15}$$

However, the optimal value is obtained by using the hopping technique [21,22], under which the signal is sampled alternatingly at $\phi_0 = \pi/2$ and $\phi_0 = -\pi/2$, and subtracting the signals from each other. When this technique is used, the MMFS for the VM is reduced by a factor of $\sqrt{2}$ compared to its value obtained by making measurements at one side only (i.e., $\phi_0 = \pm\pi/2$). The MMFS due to the VM is reduced further by another factor of $\sqrt{2}$ if one monitors the signals in both output ports of the UMZI. Finally, if the measurement is carried out $M \equiv \tau_M/\tau_D$ times, the frequency uncertainty is reduced by a factor of $\sqrt{M}$ . Combining these effects, we get:

$$\Delta\omega_{MIN,VM} = \frac{1}{\sqrt{2S_M}} \cdot \frac{1}{2\tau_{STL}} \cdot \frac{e^X}{X}; \quad X \equiv \frac{\tau_D}{2\tau_{STL}}; \quad S_M \equiv S_0\,\tau_M/\tau_D = P_{out}\tau_M/\hbar\omega_o \qquad (16)$$

Consider first the limit where $\tau_D \ll \tau_{STL}$ so that $X \to 0$, with $2\tau_{STL}X = \tau_D = 1/\gamma_S$. After some simple algebra, we then get:

$$\Delta\omega_{MIN,VM} = \frac{\gamma_S}{\sqrt{2S_M}} = \varepsilon_0\Gamma_0 \qquad (17)$$

which agrees exactly with the result found in Eqn. (5), derived using general arguments based on uncertainty relations only.

Inserting the results from Eqns. (14) and (17) into Eqn. (11), we find that the MMFS due to the effect of both the SE and the VM is given by:

$$\Delta\omega_{MIN} = \sqrt{\Delta\omega_{MIN,SE}^2 + \Delta\omega_{MIN,VM}^2} = \sqrt{\Gamma_0^2 + \left(\varepsilon_0\Gamma_0\right)^2} = \Gamma_0\sqrt{1+\varepsilon_0^2} \qquad (18)$$

As an explicit example, consider a case where $\tau_{STL} = 1\,\text{sec}$, $\tau_{LC} = 10^{-6}\,\text{sec}$, and the path length difference for the UMZI is 0.3 m, so that $\tau_D = 10^{-9}\,\text{sec}$. We then have $X = \tau_D/2\tau_{STL} = 0.5*10^{-9}$ and $\varepsilon_o = \gamma_S/\gamma_{LC} = \tau_{LC}/\tau_D = 10^3$. Thus, for such an UMZI, the MMFS would be dominated by the effect of the VM rather than SE, with a value of $\sim 10^3\,\Gamma_0$, according to Eqn. (18), in sharp violation of the common perception that the value of the MMFS is $\Gamma_0$.

For the more general case, it is easy to see that the minimum value of this quantity occurs at $X=1$, which corresponds to $\tau_D = 2\tau_{STL}$. For this choice of the decay time, we get:

$$\Delta\omega_{MIN,VM} = \frac{e}{4}\frac{1}{\langle n\rangle_{LC}}\Gamma_0 = e\varepsilon_o\Gamma_0 \qquad (19)$$

Since the number of photons inside the laser cavity, $\langle n\rangle_{LC}$ , is very large for most lasers, we see that for this optimal choice of the delay time, the MMFS due to the VM becomes negligible compared to $\Gamma_0$. For the example considered above, we have $\langle n\rangle_{LC} = \tau_{STL}/2\tau_{LC} = 0.5*10^6$, and $\varepsilon_o = \tau_{LC}/\tau_D = \tau_{LC}/2\tau_{STL} = 0.5*10^{-6}$, so that $\Delta\omega_{MIN,VM} \simeq 1.4*10^{-6}\Gamma_0$.

Since the value of $\Delta\omega_{MIN,VM}$ is so small, the constraint for Eqn. (8) to be valid (i.e., the slope of the signal must be essentially constant over the frequency span corresponding to the frequency fluctuation) is easily satisfied. However, this is not the case for the MMFS due to SE; as such, Eqns. (9) and (10) can no longer be used for determining the value of $\Delta\omega_{MIN,SE}$. In the Appendix, we use a general approach to determine the values of both $\Delta\omega_{MIN,VM}$ and $\Delta\omega_{MIN,SE}$. This analysis agrees with the value of $\Delta\omega_{MIN,VM}$ shown in Eqn. (19). It also shows that the value of $\Delta\omega_{MIN,SE}$ remains close to $\Gamma_0$, within a factor of $e/2 \simeq 1.36$ , determined by the phase diffusion caused by the SE only. It should be noted that the path length difference for the UMZI in this case would be $6*10^8$ meters, which is highly impractical. Thus, in general, for an UMZI, operating under this optimal condition is very difficult. One possible technique that may make it possible to reach this limit in practice is to make use of slow-light [23,24], as described in ref. [25].

### 3.2 Minimizing the MMFS for a Laser Employing a Fabry Perot Cavity

Next, we consider the case where a Fabry-Perot cavity (FPC) is used to measure the frequency shift of a laser. For simplicity of analysis, we consider an FPC configured as a ring resonator, illustrated schematically in Figure 3. Briefly, we assume that all the power exiting the test laser is sent into the FPC, consisting of two partial reflectors and a perfect reflector. The cavity length is tuned to be resonant with the nominal frequency of the laser, by monitoring the transmitted light. A shift in the laser frequency would produce a corresponding change in the output, deviating from its peak value. Measurement of this change can then be used to infer the change in the shift in the laser frequency. We denote by $\gamma_{co}$ the width of the cavity resonance, so that the cavity decay time is $\tau_{co} = 1/\gamma_{co}$ .

Consider first the conventional model which does not take into account the effect of the finite value of the STL. As we will show later, the result found when ignoring the STL is valid when the width of the FPC resonance is much broader than the STL (i.e., $\gamma_{co} >> \gamma_{STL}$ ). In this limit, the output signal can be expressed as $S = S_0\left[\gamma_{co}^2/\left(\gamma_{co}^2 + \tilde{\omega}^2\right)\right]$, where $\tilde{\omega}$ represents the shift in the laser frequency away from the resonance in the FPC. As such, the spectral width of the measurement system for the FPC under this condition is $\gamma_S = \gamma_{co}$, so that $\tau_S = 1/\gamma_S = \tau_{co}$

The uncertainty in the frequency of the laser due to the VM is again given, according to Eqn. (8), by $\Delta\omega_{VM} = \left|\Delta S/(\partial S/\partial\tilde{\omega})\right|$, where $\Delta S = \sqrt{S}$ is the shot-noise in the signal. Using the hopping technique [21,22], which samples the signals at $\tilde{\omega} = \pm\gamma_{co}/\sqrt{2}$ for optimal sensitivity, we find that, for a single measurement time of $\tau_S = 1/\gamma_S$ the MMFS is given by:

$$\Delta\omega_{MIN,VM} \simeq \gamma_S\sqrt{\frac{\gamma_S\hbar\omega_0}{2P_{out}}} = \frac{\gamma_S}{\gamma_{LC}}\sqrt{\gamma_S\gamma_{STL}} = \varepsilon_o\sqrt{\gamma_S\gamma_{STL}} \tag{20}$$

where $P_{out}$ is the output power of the laser. If the measurement is carried out $M \equiv \tau_M/\tau_S$ times, the frequency uncertainty is reduced by a factor of $\sqrt{M}$ , so that:

$$\Delta\omega_{MIN,VM} \simeq \varepsilon_o\sqrt{\gamma_m\gamma_{STL}} = \varepsilon_o\Gamma_0 \tag{21}$$

in agreement with Eqn. (5) and Eqn. (17).

We note that Dorschner et al. [13] finds essentially the same value of the MMFS for a laser when using the FPC, using the argument based on the Heisenberg uncertainty relation, $\Delta\phi \cdot \Delta n = 1/2$, which is briefly summarized here [26]. For a single measurement, the number of photons is given by $n_c = P_{out}\tau_{co} / \hbar\omega_0$, the number of photons in the FPC cavity (which is *not* the same as the number of photons in the laser cavity). Noting that $\Delta n = \sqrt{n}$, and that the phase has to be measured twice to determine the frequency, yields $\Delta\omega = 1\big/\left[\tau_{co}\sqrt{2n_c}\right]$. When the measurement is carried out for a total time of $\tau_m$, the total number of photons becomes $n_c\left(\tau_m/\tau_{co}\right)$, which implies an increase in the number uncertainty and a decrease in the phase uncertainty by a factor of $\sqrt{\left(\tau_m/\tau_{co}\right)}$, which in turn reduces the frequency uncertainty by the same factor, yielding the same result as Eqn. (21). Ezekiel and Balsamo [17] also found essentially the same result, but employing the shot-noise based argument we have presented above. It should be noted that the result found by Ezekiel and Balsamo also agreed closely with experimental data presented in ref. [17], since the constraint for validity (i.e., $\gamma_{co} >> \gamma_{STL}$) was clearly satisfied.

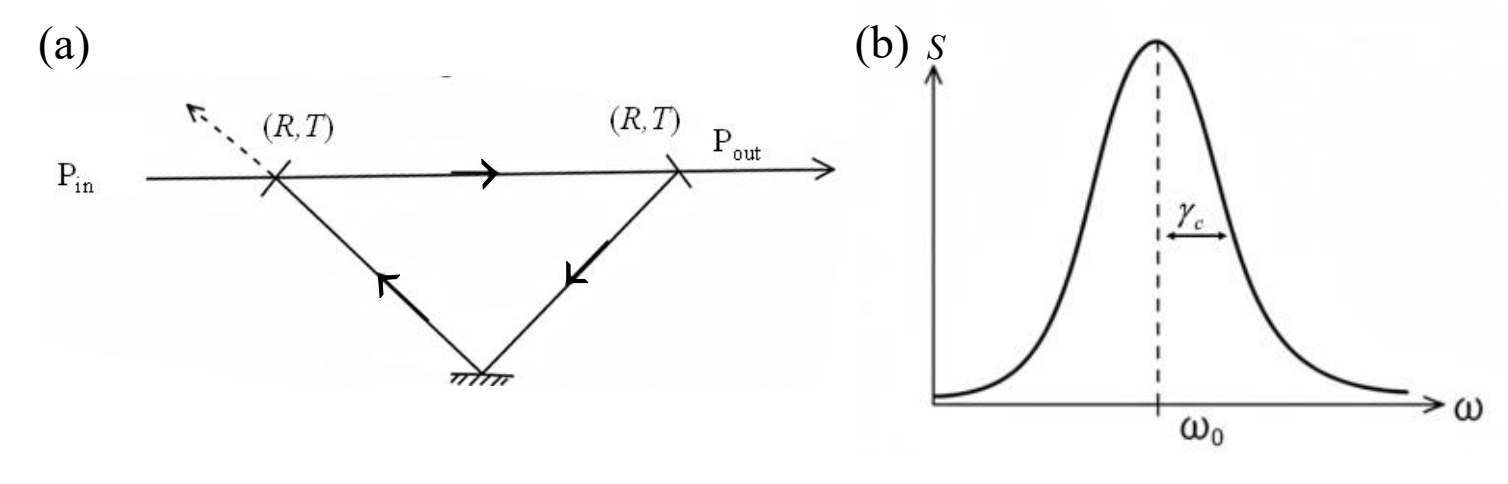

Figure 3: (a). Schematic for a FPC. (b) Transmission function for the FPC.

It is instructive to compare the MMFS shown in Eqn. (2), claimed by Dorschner et al. in [13] to be the standard limit, with the MMFS obtained using an FPC, shown in Eqn. (20), also derived by Dorschner et al. in the same paper. Specifically, we see that $\Delta\omega_{MIN,VM} = \left(\gamma_{co}/\gamma_{LC}\right)\sqrt{\gamma_M\gamma_{STL}} = \left(\gamma_{co}/\gamma_{LC}\right)\Gamma_0$. Thus, it is obvious that by making an FPC with a linewidth much narrower than the linewidth of the cavity used for making the laser, it is possible to obtain an MMFS much smaller than what Dorschner et al. claimed to be the standard limit. It is curious that Dorschner et al. did not make this comparison in their paper [13], which would have shown clearly the invalidity of their claim about the standard limit. The reason is that they implicitly assumed that the quality factor of both the laser and the FPC are the same (so that $\gamma_{co}/\gamma_{LC} = 1$), which clearly does not have to be the case. Because of this assumption, they claimed that the limit for the FPC found in Eqn. 9 of ref. [13] is the same as the standard limit for the laser found in Eqn. (12) of the same reference.

Of course, the MMFS shown in Eqns. (20) is not complete when the effect of SE is taken into account, for two reasons. First, SE causes the STL to have a finite value, which may or may not be negligible. As we describe below, the signal measured at the output of the FPC gets modified when the effect of the STL is taken into account. As a result, the shot-noise is affected, thereby changing the value of the MMFS due to the shot-noise. Second, the direct contribution of the phase diffusion due to SE, as shown in Eqn. (2), must also be taken into account, and this would become the dominant contributor to the MMFS in the limit of $\gamma_{co} << \gamma_{LC}$. In what follows, we first derive the correct expression for the value of $\Delta\omega_{MIN,VM}$, and identify conditions under which the value of $\Delta\omega_{MIN,VM}$ is much less than $\Delta\omega_{MIN,SE}$, so that the correct total MMFS, which is given by Eqn. (7), becomes approximately $\Delta\omega_{MIN,SE} = \sqrt{\gamma_M \gamma_{STL}} = \Gamma_0$.

We denote by $L$ the length of the cavity, so that the cavity roundtrip time is $\tau_{RT} = L / c_0$. The signal due to the interference between the *n*-th and *m*-th round trips is attenuated by the factor of $\exp\left(-|m-n|\tau_{RT} / 2\tau_{STL}\right)$, as shown in [27]. When this factor is taken into account while summing over infinite number of round trips, the signal at the output of the FPC becomes [27]:

$$S = S_0 \frac{\gamma_c^2 a(\Delta)}{\gamma_c^2 + \tilde{\omega}^2}; \quad \gamma_c = \frac{2c_0}{L} \cdot \frac{1}{\sqrt{b(\Delta)}}; \quad a(\Delta) = \frac{(1-R)^2 (1-\Delta^2 R^2)}{(1-R^2)(1-\Delta R)^2}; \quad b(\Delta) = \frac{4\Delta R}{(1-\Delta R)^2} \quad (22)$$

where $\Delta \equiv \exp(-\tau_{RT} / 2\tau_{STL})$. It is easy to see that in the limit of a vanishingly small STL, which corresponds to $\Delta = 1$, we recover the conventional expression for this signal. Using this expression, and for the optimum operating point used in the hopping technique, we have shown [28] that the VM contribution to the MMFS is given by [29]:

$$\Delta\omega_{MIN,VM} \simeq \beta\xi \frac{\gamma_S}{\gamma_{LC}} \sqrt{\gamma_{STL}\gamma_m} = \beta\xi\varepsilon_o\Gamma_0 \quad (23)$$

where $\beta = 3\sqrt{3}/4 \simeq 1.3$, $\gamma_S = \gamma_{co} = c_0(1-R)/L\sqrt{R}$, and the STL-dependent correction factor, $\xi$, is given by:

$$\xi = \left(\frac{1-\Delta R}{1-R}\right)^2 \sqrt{\frac{1-R^2}{\Delta(1-\Delta^2 R^2)}} \quad (24)$$

The value of $\gamma_c$ represents the width of the cavity resonance when taking into account the effect of the STL. For the limiting case where the STL is vanishingly small, corresponding to $\Delta = 1$, we get $\xi = 1$ and recover Eqn. (20) under the approximation of $\beta \simeq 1$. From Eqns. (23) and (24), it is not obvious whether there is an optimum value of $\Delta$ for which the MMFS due to the VM is minimized. In what follows, we address this question.

We define $2\rho \equiv \tau_{co} / \tau_{STL} = \tau_S / \tau_{STL} = \gamma_{STL} / \gamma_S$ as the ratio between the empty-cavity decay time for the FPC and the coherence time of the laser, so that:

$$\varepsilon_o = \frac{\gamma_S}{\gamma_{LC}} = \frac{\gamma_S}{\gamma_{STL}} \cdot \frac{\gamma_{STL}}{\gamma_{LC}} = \frac{1}{2\rho} \cdot \frac{1}{2\langle n \rangle_{LC}} \quad (25)$$

We now rewrite Eqn. (23) as:

$$\Delta\omega_{MIN,VM} = \frac{\beta}{2}\cdot\frac{1}{2\langle n\rangle_{LC}}\cdot\Gamma_0\cdot\frac{1}{\rho}\left(\frac{1-\Delta R}{1-R}\right)^2\sqrt{\frac{1-R^2}{\Delta\left(1-\Delta^2R^2\right)}} \equiv \frac{\beta}{2}\cdot\frac{\Gamma_0}{2\langle n\rangle_{LC}}\cdot B(\rho,R) \tag{26}$$

where the STL dependent function, $B(\rho,R)$ is given by:

$$B(\rho,R) = \frac{1}{\rho}\sqrt{\frac{1+R}{\Delta(1+\Delta R)}}\left(\frac{1-\Delta R}{1-R}\right)^{3/2} \equiv \frac{1}{\rho}\sqrt{\frac{1+R}{\Delta(1+\Delta R)}}g(\rho,R)^{3/2} \tag{27}$$

where $g(\rho,R)=(1-\Delta R)/(1-R)$. Next, we note that the dephasing parameter, $\Delta$, can be expressed as $\Delta=\exp\left[-\rho(1-R)/\sqrt{R}\right]$. The fundamental width of the FPC resonator, $\gamma_{co}$, becomes vanishingly small when $R \to 1$. As such, it is reasonable to expect that the smallest value of the VM contribution to the MMFS would occur in this limit. For $R \to 1$, we get $\Delta|_{R\to1} \simeq 1-\rho(1-R)/\sqrt{R}$. Therefore, we can express $g(\rho,R)$ and $B(\rho,R)$ as:

$$g|_{\rho,R\to1} \approx \left[1-\left(1-\rho\frac{1-R}{\sqrt{R}}\right)R\right]/(1-R) = 1+\rho\sqrt{R} \simeq 1+\rho \tag{28}$$

$$B|_{\rho,R\to1} = (1+\rho)^{3/2}/\rho \tag{29}$$

Taking the derivative of $B$ with respect to $\rho$ and making it equal to 0, we find that minimum of $B$ occurs at $\rho=2$, for $R \to 1$. The corresponding value of $B$ is $\sim 3\sqrt{3}/2 \simeq 2.6$.

Figure 4(a) shows the numerically computed value of $B$ as a function of $R$ for $\rho=2$. As can be seen, the minimum value of $B$ estimated above analytically agrees with the numerical

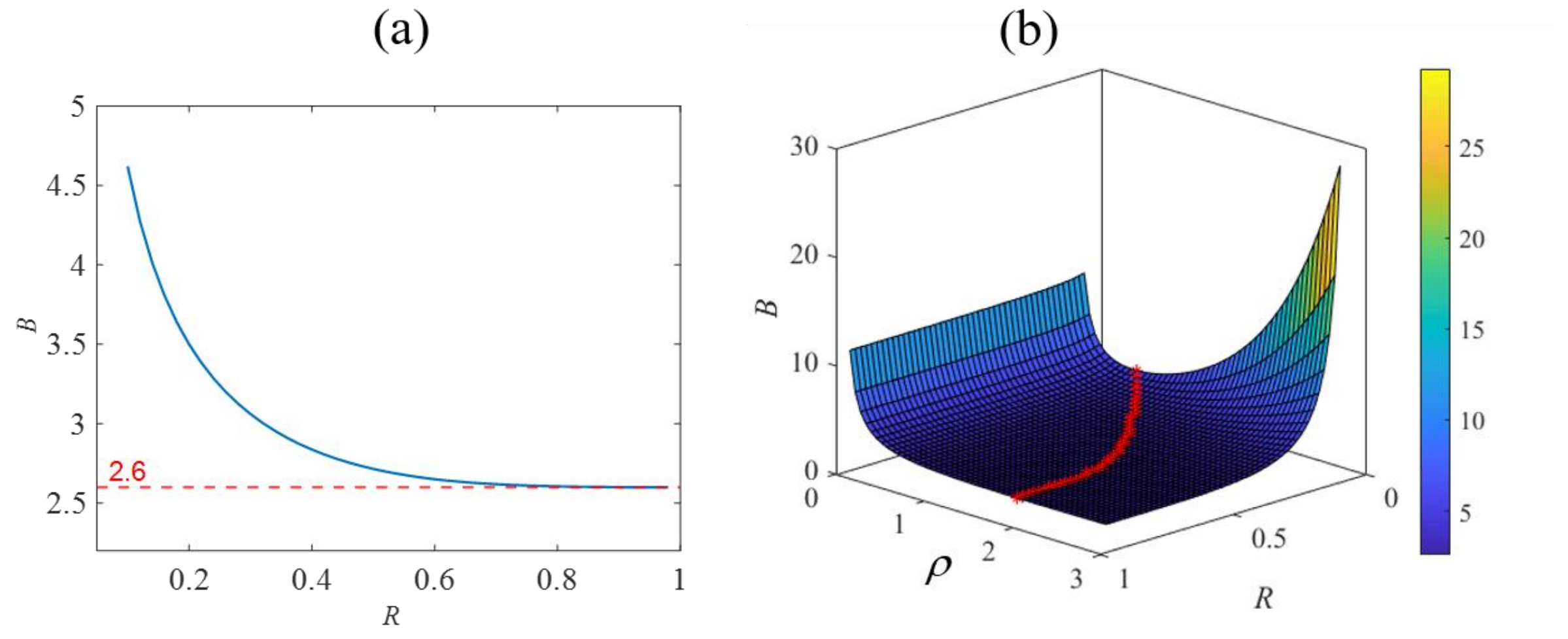

Figure 4: (a) The numerical value of B as a function of R, for $\rho=2$. As can be seen, it agrees with the estimated value of ~2.6 as R approaches unity. (b) Plot of B when both $\rho$ and R are varied

evaluation as $R \to 1$. Figure 4(b) shows a plot of $B$ when both $\rho$ and $R$ are varied. The red asterisks indicate the minimum value of $B$ for a specific value of $R$. The overall minimum value for $B$ occurs at $\rho = 2$ as $R \to 1$. Therefore, the value $B\big|_{\rho=2,R\to 1} \simeq 2.6$ is the global minimum. From Eqn. (26), it then follows that the global minimum for the VM contribution to the MMFS for the FPC is:

$$\Delta\omega_{MIN,VM} \simeq 0.85 \frac{1}{\langle n \rangle_{LC}} \Gamma_0 = 6.8\varepsilon_o \Gamma_0 \tag{30}$$

which occurs for $\rho = 2$, so that $\varepsilon_o = 1/8\langle n \rangle_{LC}$, according to Eqn. (25). For the example laser considered above, we have $\langle n \rangle_{LC} = \tau_{STL}/2\tau_{LC} = 0.5*10^6$, so that $\Delta\omega_{MIN,VM} \simeq 1.7*10^{-6}\Gamma_0$, which is negligibly small compared to $\Gamma_0$.

It should be noted that this value of $\rho$ corresponds to the cavity decay time being equal to four times the coherence time of the laser: $\tau_{co} = \tau_S = 4\tau_{STL}$. In a qualitative sense, this decay time is the time duration of the copy of the beam that has undergone roughly the highest useful number of bounces. In comparison, the time duration of the beam that has traversed the cavity only once is negligibly small. Thus, the mean time for a single measurement is the average of these two, equaling $2\tau_{STL}$. The fact that the global minimum of the VM contribution to the MMFS occurs under this condition is again consistent with the conclusion reached for the UMZI that a time duration close to $\tau_{STL}$ is the optimal duration of the single measurement time.

Since the value of $\Delta\omega_{MIN,VM}$ is so small, the constraint for Eqn. (8) to be valid (i.e., the slope of the signal must be essentially constant over the frequency span corresponding to the frequency fluctuation) is easily satisfied. However, this is not the case for the MMFS due to SE, since $\tau_S$ is now comparable to $\tau_{STL}$; as such, Eqns. (9) and (10) can no longer be used for determining the value of $\Delta\omega_{MIN,SE}$ very accurately. What is needed is an analysis similar to the one shown in the Appendix for the UMZI. Unfortunately, such an analysis is analytically intractable due to the complicated form of the expression for the FPC signal shown in Eqn. (22). However, under very similar conditions, we showed for the case of the UMZI that the value of $\Delta\omega_{MIN,SE}$ remains close to $\Gamma_0$, within a factor of $e/2 \simeq 1.36$, even when $\tau_S = 2\tau_{STL}$. As such, it is reasonable to conclude that for the FPC, the value of $\Delta\omega_{MIN,SE}$ remains close to $\Gamma_0$ when the value of $\Delta\omega_{MIN,VM}$ is minimized. Thus, under optimal operating conditions, the overall MMFS for the FPC is given approximately by $\Delta\omega_{MIN,SE} \simeq \Gamma_0$. It should also be noted that this is also the limit for measuring the change in the resonance frequency of an FPC.

As an explicit example, consider an FPC, with a mirror reflectivity of $R = 0.99998$, which is typical of the state of the art for high quality mirrors. To achieve the minimized MMFS corresponding to $\rho = 2$ for the example laser considered above would require a perimeter length of $\sim 2.42*10^4$ meters, which also too large to realize experimentally on earth. However, we have recently shown [28] that it may be possible to make use of slow light to reduce the length needed for an FPC.

### 3.3 Minimizing the MMFS for a Laser via Heterodyning with a Reference Laser

We consider next the technique of measuring the frequency shift of a test laser by heterodyning with a reference laser. This process is illustrated schematically in Figure 5. The outputs of the two lasers are combined with a 50:50 beam-splitter, and observed with a photodetector (PD). The signal from the PD is fed into a spectral analyzer (SA). The spectrum of the beat signal is centered at the frequency corresponding to the difference in the frequencies of the reference laser and the test laser. Assuming that the frequency of the reference laser is perfectly stable, measurement of the shift in the peak of the beat-signal enables one to measure the frequency shift of the test laser.

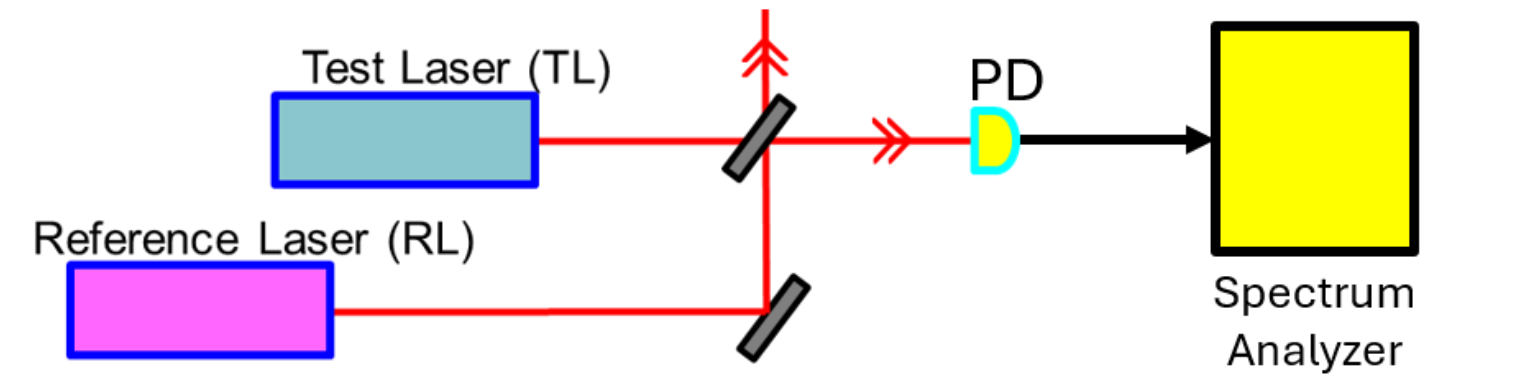

Figure 5: Schematic illustration of the technique of heterodyne detection for measuring the frequency shift of a test laser.

We assume that the power of the reference laser is much larger than that of the test laser. Furthermore, we assume that the STL of the reference laser is much less than that of the test laser. Under these conditions, it has been shown [30] that the effective signal-to-noise ratio in the beat signal is given by the power and fluctuation thereof of the test laser only.

The effective heterodyne beat signal observed on the SA can be expressed as:

$$S = \frac{S_0}{2}\frac{A\tilde{\gamma}^2}{\tilde{\gamma}^2 + \tilde{\omega}^2}. \tag{31}$$

Here, $\tilde{\omega}$ is defined as the deviation from the quiescent value of the laser frequency, $S_o = P_{out}\tau_S / \hbar\omega_o$ represents the number of photons emerging from the test laser during the single measurement time, $\tau_S$, and $\omega_o$ is the nominal frequency of the test laser. In this expression, we have also defined a factor that accounts for the finite STL of the test laser: $A \equiv \exp[-\tau_{SA}/2\tau_{STL}] = \exp[-\gamma_{STL}/2\gamma_{SA}]$, where $\gamma_{SA} = 1/\tau_{SA}$ is the bandwidth of the spectrum analyzer. As such, $\gamma_{SA}$ is the spectral width of the measurement system, $\gamma_S = 1/\tau_S$, which also means that $\tau_S = \tau_{SA}$. We have also defined the width of the beat spectrum as $\tilde{\gamma} = [\gamma_{STL}^2 + \gamma_{SA}^2]^{1/2}$.

The MMFS of the test laser due to the VM is again given by $\Delta\omega_{VM} = |\Delta S/(\partial S/\partial\tilde{\omega})|$, where $\Delta S = \sqrt{S}$ is the shot-noise in the signal. If the measurement is made at $\tilde{\omega} = \pm\tilde{\gamma}$, we get:

$$\Delta\omega_{MIN,VM} = \frac{2\gamma_{STL}}{\sqrt{S_o}} f(X) = \frac{\sqrt{2}}{\langle n\rangle_{LC}}\sqrt{\gamma_S\gamma_{STL}} f(X); \quad f(X) = e^{X/2}\left(1+1/4X^2\right)^{1/2}; \quad X \equiv \tau_S/2\tau_{STL} \tag{32}$$

If the hopping method is used at the optimal operating point at $\tilde{\omega} = \pm\tilde{\gamma}/\sqrt{2}$, the uncertainty is reduced by a factor of ~1.54. If the measurement is carried out in both output ports of the beamsplitter that combines the test laser with the reference laser, the uncertainty is reduced by another factor of $\sqrt{2}$. Furthermore, if the measurement is carried out $M \equiv \tau_M / \tau_S$ times, the frequency uncertainty is reduced by a factor of $\sqrt{M}$. Taking these into account, we get:

$$\Delta\omega_{MIN,VM} \simeq \frac{0.65}{\langle n \rangle_{LC}} \sqrt{\gamma_M \gamma_{STL}} f(X) = \frac{0.65}{\langle n \rangle_{LC}} f(X) \Gamma_0 \tag{33}$$

Consider first the limit where the bandwidth of the spectrum analyzer is much larger than the STL, which corresponds to $X << 1$. In this limit, we have $f(X) \simeq 1/2X = \gamma_S / \gamma_{STL}$, so that, after some algebra, we get:

$$\Delta\omega_{MIN,VM} \simeq 1.3 \varepsilon_o \Gamma_0; \quad \varepsilon_o = \gamma_S / \gamma_{LC} \tag{34}$$

which is in close agreement with Eqn. (5) derived from general uncertainty relations, within a factor of ~1.3. In this limit, the MMFS due to SE would be $\Delta\omega_{MIN,SE} = \Gamma_0$. As such, the overall MMFS would be much larger than $\Gamma_0$ if the spectral analyzer bandwidth is much larger than the linewidth of the test laser cavity: $\gamma_S >> \gamma_{LC}$.

Setting $\partial f / \partial X = 0$, we find that the smallest value of the MMFS in Eqn. (33) occurs at $X \simeq 0.69$, which corresponds to $\tau_S = 1.38 \tau_{STL}$. For this value of $X$, we find:

$$\Delta\omega_{MIN,VM} \simeq \frac{1.13}{\langle n \rangle_{LC}} \Gamma_0 \simeq 3.12 \varepsilon_o \Gamma_0 \tag{35}$$

For the example laser considered above, we have $\langle n \rangle_{LC} = \tau_{STL} / 2\tau_{LC} = 0.5 * 10^6$, so that $\Delta\omega_{MIN,VM} \simeq 2.3 * 10^{-6} \Gamma_0$, which is negligibly small compared to $\Gamma_0$. The fact that the global minimum of the VM contribution to the MMFS occurs under this condition is again consistent with the conclusion reached for the UMZI that a time duration close to $\tau_{STL}$ is the optimal duration of the single measurement time.

Since the value of $\Delta\omega_{MIN,VM}$ is so small, the constraint for Eqn. (8) to be valid (i.e., the slope of the signal must be essentially constant over the frequency span corresponding to the frequency fluctuation) is easily satisfied. However, this is not the case for the MMFS due to SE, since $\tau_S$ is now comparable to $\tau_{STL}$; as such, Eqns. (9) and (10) can no longer be used for determining the value of $\Delta\omega_{MIN,SE}$ very accurately. What is needed is an analysis similar to the one shown in the Appendix for the UMZI. To see the challenge in carrying out such an analysis in this case, we note first that the signal expressed in Eqn. (31) is an effective one, which includes the broadening due to the spectrum analyzer modeled as a classical device, and does not explicitly take into account the presence of the reference laser. As such, it is not obvious how to map this expression into an effective superposition of coherent states with randomly varying phases, which is needed for the type of model used in the Appendix. However, under very similar conditions, we showed for the case of the UMZI that the value of $\Delta\omega_{MIN,SE}$ remains close to $\Gamma_0$, within a factor of $e/2 \simeq 1.36$, even when $\tau_S = 2\tau_{STL}$. As such, it is reasonable to conclude that for the heterodyning scheme, the value of $\Delta\omega_{MIN,SE}$ remains close to $\Gamma_0$ when the value of $\Delta\omega_{MIN,VM}$ is minimized. Thus, under optimal operating conditions, the overall MMFS for the heterodyning scheme is given approximately by $\Delta\omega_{MIN,SE} \simeq \Gamma_0$.

## 4. Summary and Conclusion

It is generally accepted that the minimum measurable frequency shift (MMFS) of a single-mode ideal laser is given by the geometric mean of the measurement bandwidth and the Schwalow-Townes Linewidth. In this paper, we show that this is incorrect in general. Specifically, we show that this result is valid only when considering the phase diffusion caused by SE, and the effect of the uncertainty in the laser frequency caused by the VM is ignored. For the general case, we show that the MMFS for such a laser is determined by a combination of phase diffusion caused by SE and the shot noise caused by the VM. For all practical sensors, the MMFS is found to be given by the geometric mean of the measurement bandwidth and the Schwalow-Townes Linewidth, multiplied by a factor which can be much larger than unity under certain conditions. We determine the optimal value for the MMFS for three different sensing modalities, an unbalanced Mach-Zehnder Interferometer (which is equivalent to a unbalanced Michelson Interferometer), a passive Fabry-Perot cavity (FPC), and heterodyning with a reference laser, and identify the conditions needed for reaching the optimal value.

**Appendix: Explicit Model of the Quantum State, Fisher Information and the Cramer-Rao Bound for the Unbalanced MZI**

Here, we consider an explicit model of the quantum state used for measuring the frequency shift. Specifically, we consider the case where the UMZI is used for this purpose. A plausible model for the quantum state at one of the output ports of the UMZI is as follows:

$$\rho = \int d\phi \mathcal{N}\left(\phi;\phi_0,\sqrt{\frac{\tau}{\tau_c}}\right)\left|\sqrt{\lambda}\right\rangle\left\langle\sqrt{\lambda}\right| \tag{36}$$

where $\phi_0$ is the phase shift resulting from the length difference in the optical paths, $\phi$ is the actual phase shift subject to phase diffusion, $\mathcal{N}\left(\phi;\phi_0,\sqrt{\tau/\tau_c}\right)$ is a normal distribution with respect to $\phi$ with the mean and standard deviation being $\phi_0$ and $\sqrt{\tau/\tau_c}$ respectively, with $\tau_c = \tau_{STL}$ being the coherence time of the laser, $\left|\sqrt{\lambda}\right\rangle$ is an eigenstate of the annihilation operator (coherent state) with the eigenvalue being $\sqrt{\lambda}$, and $\lambda = N\left(1+\cos\phi\right)/2$. The probability of detecting $n$ photons is given by

$$\langle n|\rho|n\rangle = \int d\phi \mathcal{N}\left(\phi;\phi_0,\sqrt{\frac{\tau}{\tau_c}}\right)\left|\left\langle n\middle|\sqrt{\lambda}\right\rangle\right|^2 = \int d\phi \mathcal{N}\left(\phi;\phi_0,\sqrt{\frac{\tau}{\tau_c}}\right)\mathcal{P}(n;\lambda) \tag{37}$$

where $\mathcal{P}(n;\lambda)$ is the Poisson distribution. In the limit $\tau/\tau_c \to 0$, the normal distribution can be approximated by a Dirac delta function, and $\langle n|\rho|n\rangle$ approximately follows a Poisson distribution.

The measured variance of $\phi$ can be calculated with the error propagation formula $\mathrm{var}(\phi_0) = \mathrm{var}(n)/\left(d\langle n\rangle/d\phi_0\right)^2$. The measured expectation value of $n$ is given by

$$\langle n\rangle = \sum_{n=0}^{\infty} n\int d\phi \mathcal{N}\left(\phi;\phi_0,\sqrt{\frac{\tau}{\tau_c}}\right)\left|\left\langle n\middle|\sqrt{\lambda}\right\rangle\right|^2 = \int d\phi \mathcal{N}\left(\phi;\phi_0,\sqrt{\frac{\tau}{\tau_c}}\right)\sum_{n=0}^{\infty} n\mathcal{P}(n;\lambda) \equiv \left\langle\langle n\rangle_n\right\rangle_\phi \tag{38}$$

where $\langle n\rangle_n$ is the expectation value of $n$ for a given $\phi$, so that $\langle n\rangle_n = \sum_{n=0}^{\infty} n\mathcal{P}(n;\lambda) = \lambda$, and $\left\langle\langle n\rangle_n\right\rangle_\phi$ is the expectation value of $\langle n\rangle_n$ over $\phi$, so that

$$\left\langle\langle n\rangle_n\right\rangle_\phi = \langle\lambda\rangle_\phi = N\langle 1+\cos\phi\rangle_\phi/2 = N\left(1+e^{-\tau/2\tau_c}\cos\phi_0\right)/2 = \langle n\rangle \tag{39}$$

The variance of $n$ can be calculated as follows:

$$\text{var}(n) \equiv \left\langle \langle n^2 \rangle_n \right\rangle_\phi - \langle n \rangle^2 = \left\langle \langle n^2 \rangle_n - \langle n \rangle_n^2 \right\rangle_\phi + \left\langle \langle n \rangle_n^2 \right\rangle_\phi - \langle n \rangle^2 = \left\langle \text{var}\left(\langle n \rangle_n\right) \right\rangle_\phi + \text{var}\left(\langle n \rangle_n\right) \quad (40)$$

Noting that for a Poisson distribution, $\text{var}\left(\langle n \rangle_n\right) = \lambda$, the first term in the final expression in Eqn. (40) can be expressed as:

$$\left\langle \text{var}\left(\langle n \rangle_n\right) \right\rangle_\phi = \langle \lambda \rangle_\phi = N\left(1 + e^{-\tau/2\tau_c} \cos\phi_0\right)/2 \quad (41)$$

Noting that $\lambda = N\left(1+\cos\phi\right)/2$ , the second term in the final expression in Eqn. (40) can be expressed as:

$$\begin{aligned} \text{var}\left(\langle n \rangle_n\right) &= \text{var}(\lambda) = \frac{N^2}{4}\text{var}(1+\cos\phi) = \frac{N^2}{4}\left[\left\langle (1+\cos\phi)^2 \right\rangle_\phi - \langle 1+\cos\phi \rangle_\phi^2\right] \\ &= \frac{N^2}{8}\left(1-e^{-\tau/\tau_c}\right)\left(1-e^{-\tau/\tau_c}\cos 2\phi_0\right) \end{aligned} \quad (42)$$

Accordingly,

$$\begin{aligned} \text{var}(\phi_0) &= \frac{\text{var}(n)}{\left(d\langle n \rangle / d\phi_0\right)^2} \\ &= \left[\frac{2\left(1+e^{-\tau/2\tau_c}\cos\phi_0\right)}{Ne^{-\tau/\tau_c}\left(\sin\phi_0\right)^2}\right] + \left[\frac{\left(1-e^{-\tau/\tau_c}\right)\left(1-e^{-\tau/\tau_c}\cos 2\phi_0\right)}{2e^{-\tau/\tau_c}\left(\sin\phi_0\right)^2}\right] \end{aligned} \quad (43)$$

The corresponding variance in the frequency can be expressed as:

$$\text{var}(\omega) = \text{var}(\phi_0)/\tau^2 = \left[\frac{2\left(1+e^{-\tau/2\tau_c}\cos\phi_0\right)}{N\tau^2 e^{-\tau/\tau_c}\left(\sin\phi_0\right)^2}\right] + \left[\frac{\left(1-e^{-\tau/\tau_c}\right)\left(1-e^{-\tau/\tau_c}\cos 2\phi_0\right)}{2\tau^2 e^{-\tau/\tau_c}\left(\sin\phi_0\right)^2}\right] \quad (44)$$

In this expression, the term in the first square bracket results from the component of $\text{var}(n)$ shown in Eqn. (41). Noting that this term is proportional to $(1/N)$, it is clear that it represents the effect of shot-noise. The term in the second square bracket results from the component of $\text{var}(n)$ shown in Eqn. (42). Noting that this term is independent of $N$, it is clear that it represents the effect of the phase diffusion only. Thus, we can write:

$$\Delta\omega_{MIN,VM} = \left[\frac{2\left(1+e^{-\tau/2\tau_c}\cos\phi_0\right)}{N\tau^2 e^{-\tau/\tau_c}\left(\sin\phi_0\right)^2}\right]^{1/2} \quad (45)$$

$$\Delta\omega_{MIN,SE} = \left[ \frac{\left(1-e^{-\tau/\tau_c}\right)\left(1-e^{-\tau/\tau_c}\cos 2\phi_0\right)}{2\tau^2 e^{-\tau/\tau_c}\left(\sin\phi_0\right)^2} \right]^{1/2} \tag{46}$$

We now consider the Cramer-Rao Bound (CRB) for determining the MMFS for this case. Consider first a general experimental configuration measuring the uncertainty in a parameter, $\theta$. The CRB [31,32,33] states that the variance of this parameter is constrained by the relation $\Delta\theta^2 \geq 1/F(\theta)$, where $F(\theta)$ is the Fisher Information (FI). When the FI is determined via consideration of classical probability distributions, the resulting constraint is called the Classical CRB (CCRB). On the other hand, if a quantum mechanical analysis is used to determine the FI, then the resulting constraint is called the Quantum CRB (QCRB). The Classical FI (CFI) can be expressed as:

$$F_C(\theta) \equiv \left\langle \left( \frac{\partial P(B,\theta)/\partial\theta}{P(B,\theta)} \right)^2 \right\rangle \tag{47}$$

Here, $B$ is a parameter that can be measured to infer the value of $\theta$, and $P(B,\theta)$ is the normalized probability distribution.

For any given experiment, the quantum form of the FI can be found by the same expression as what is shown in Eqn. (47), provided that the probability distribution is determined from the quantum mechanical description of the process. As an example, consider a system described by a density operator, $\rho$, which has undergone a unitary transformation from $\rho_o$ of the form $\rho = \exp(-iA\theta)\rho_o \exp(iA\theta)$, where the operator $A$ represents an observable. Since any Hermitian operator can be diagonalized, the density operator can be expressed as $\rho = \sum_k \lambda_k |k\rangle\langle k|$ where $\{\lambda_k\}$ are the eigenvalues. The Quantum FI (QFI) for this system is found to be [31]:

$$F_Q[\rho, A] \equiv 2\sum_{k,l} \frac{(\lambda_k - \lambda_l)^2}{(\lambda_k + \lambda_l)} \left|\langle k|A|l\rangle\right|^2 \tag{48}$$

We now consider the application of the CRB to the process of measuring the frequency shift of a laser.

For the specific case considered here, The Fisher information can be expressed as:

$$\frac{1}{\mathrm{var}(\omega)} = \left\langle \left\{ \frac{1}{\langle n|\rho|n\rangle} \left( \frac{d\langle n|\rho|n\rangle}{d\omega} \right) \right\}^2 \right\rangle_n \tag{49}$$

This expression is not analytically tractable for quantum system being considered here. The results of numerical computation are shown in Figure 6. In the limit $N \gg 1$, the Fisher information estimated by error propagation converges to the rigorously computed Fisher information. In the limit $\tau \ll \tau_c$ and $N \gg 1$, the Fisher information approaches the value corresponding to the intrinsic variance of the phase shift.

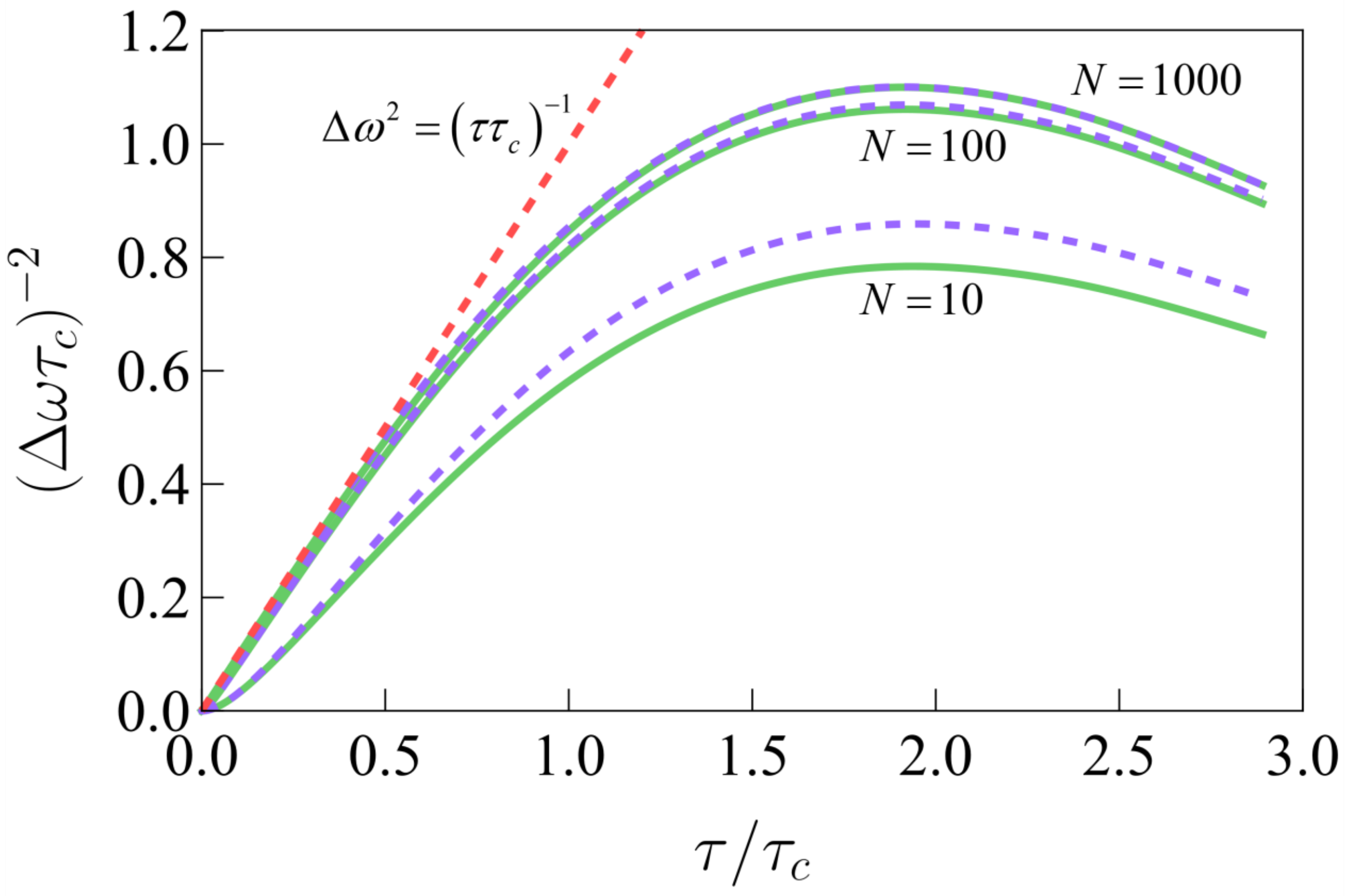

Figure 6. Fisher information rigorously calculated with the numerical method (purple dashed) and approximated by error propagation (green solid) for various numbers of photons. In the limit $\tau \ll \tau_c$ and $N \gg 1$, the Fisher information approaches the value corresponding to the intrinsic variance of the phase shift, shown by the red dashed line.

We now compare these results with the ones found earlier using explicit considerations of the effects of the SE and the VM, as documented in Sections 2 and 3. We first note the following correspondence between the notation used in Eqns . (45) and (46) and the notation used earlier in Sections 2 and 3:

$$\tau \Leftrightarrow \tau_S; \quad \tau_c \Leftrightarrow \tau_{STL}; \quad N \Leftrightarrow S_0 = P_0\tau_S/\hbar\omega_0 \tag{50}$$

Furthermore, we recall that we had defined $X \equiv \tau_S / 2\tau_{STL}$ . As such, we can write the MMFS due to the effect of the VM (i.e., shot noise) as:

$$\Delta\omega_{MIN,VM} = \left[ \frac{2\left(1+e^{-X}\cos\phi_0\right)}{S_0 \tau_S^2 e^{-2X} \left(\sin\phi_0\right)^2} \right]^{1/2} \tag{51}$$

Similarly, we can write the MMFS due the effect of the phase diffusion caused by SE as:

$$\Delta\omega_{MIN,SE} = \left[ \frac{\left(1-e^{-2X}\right)\left(1-e^{-2X}\cos 2\phi_0\right)}{2\tau_S^2 e^{-2X} \left(\sin\phi_0\right)^2} \right]^{1/2} \tag{52}$$

If the measurement is carried out $M = \tau_M / \tau_S = \gamma_S / \gamma_M$ times, the frequency uncertainty is reduced by a factor of $\sqrt{M}$ . We also recalled that we had defined $S_M = P_0 \tau_M / \hbar\omega_0$ Therefore, we can write:

$$\Delta\omega_{MIN,VM} = \left[ \frac{2\left(1+e^{-X}\cos\phi_0\right)}{S_0 \tau_S^2 e^{-2X} \left(\sin\phi_0\right)^2} \frac{\tau_S}{\tau_M} \right]^{1/2} = \left[ \frac{2\left(1+e^{-X}\cos\phi_0\right)}{S_M \tau_S^2 e^{-2X} \left(\sin\phi_0\right)^2} \right]^{1/2} \tag{53}$$

$$\Delta\omega_{MIN,SE} = \left[ \frac{\left(1-e^{-2X}\right)\left(1-e^{-2X}\cos 2\phi_0\right)}{2\tau_S^2 e^{-2X} \left(\sin\phi_0\right)^2} \frac{\tau_S}{\tau_M} \right]^{1/2} = \left[ \frac{\left(1-e^{-2X}\right)\left(1-e^{-2X}\cos 2\phi_0\right)}{2\tau_S \tau_M e^{-2X} \left(\sin\phi_0\right)^2} \right]^{1/2} \tag{54}$$

We now consider the two limits studied in Section 3, and discuss sequentially the MMFS due to the SE and that due to the VM for each limit.

*Limit 1:* $\tau_S \ll \tau_{STL}$*, corresponding to* $X \to 0$*:*

The MMFS due to the VM becomes:

$$\Delta\omega_{MIN,VM} = \left[ \frac{2\left(1+\cos\phi_0\right)}{S_M \tau_S^2 \left(\sin\phi_0\right)^2} \right]^{1/2} \tag{55}$$

We recall that when the hopping technique is used, the MMFS for the VM is reduced by a factor of $\sqrt{2}$. The MMFS due to the VM is reduced further by another factor of $\sqrt{2}$ if one monitors the signals in both output ports of the UMZI. Under these conditions, we get:

$$\Delta\omega_{MIN,VM} = \left[ \frac{\left(1+\cos\phi_0\right)}{2S_M \tau_S^2 \left(\sin\phi_0\right)^2} \right]^{1/2} \tag{56}$$

For the optimal operating point of $\phi_0 = \pm\pi/2$, we get:

$$\Delta\omega_{MIN,VM} = 1\Big/\left(\sqrt{2S_M}\tau_S\right) = \varepsilon_0\Gamma_0; \quad \varepsilon_0 = \gamma_S/\gamma_{LC}; \quad \Gamma_0 = \sqrt{\gamma_M\gamma_{STL}} \tag{57}$$

which agrees exactly with Eqn. (17). It can be shown that the agreement holds for operation at any value of $\phi_0$.

The MMFS due to the SE becomes:

$$\Delta\omega_{MIN,SE} \simeq \left[\frac{\left(1-e^{-2X}\right)}{\tau_S\tau_M}\right]^{1/2} \simeq \left[\frac{2X}{\tau_S\tau_M}\right]^{1/2} = \left[\frac{1}{\tau_{STL}\tau_M}\right]^{1/2} = \Gamma_0 \tag{58}$$

which agrees exactly with Eqn. (10). It should be noted that this result holds for any value of $\phi_0$ in this limit.

*Limit 2: $\tau_S = 2\tau_{STL}$, corresponding to $X = 1$:*

The MMFS due to the VM becomes:

$$\Delta\omega_{MIN,VM} = \left[\frac{2\left(1+e^{-1}\cos\phi_0\right)}{S_M\tau_S^2e^{-2}\left(\sin\phi_0\right)^2}\right]^{1/2} \tag{59}$$

We again recall that when the hopping technique is used, the MMFS for the VM is reduced by a factor of $\sqrt{2}$. The MMFS due to the VM is reduced further by another factor of $\sqrt{2}$ if one monitors the signals in both output ports of the UMZI. Under these conditions, we get:

$$\Delta\omega_{MIN,VM} = \left[\frac{\left(1+e^{-1}\cos\phi_0\right)}{2S_M\tau_S^2e^{-2}\left(\sin\phi_0\right)^2}\right]^{1/2} \tag{60}$$

For the optimal operating point of $\phi_0 = \pm\pi/2$, we get:

$$\Delta\omega_{MIN,VM} = e\Big/\left(\sqrt{2S_M}\tau_S\right) = e\varepsilon_0\Gamma_0 \tag{61}$$

which agrees exactly with Eqn. (19). It can be shown that the agreement holds for operation at any value of $\phi_0$. It can also be shown that, for the MMFS due to the VM, the agreement would also hold for any value of $X$ and any value of $\phi_0$.

The MMFS due to the SE becomes:

$$\Delta\omega_{MIN,SE} = \left[ \frac{\left(1-e^{-2}\right)\left(1-e^{-2}\cos 2\phi_0\right)}{2\tau_S \tau_M e^{-2} \left(\sin\phi_0\right)^2} \right]^{1/2} \tag{62}$$

For the optimal operating point of $\phi_0 = \pm\pi/2$, we get:

$$\Delta\omega_{MIN,SE} = \left[ \frac{\left(1-e^{-4}\right)}{2\tau_S \tau_M e^{-2}} \right]^{1/2} \simeq \left[ \frac{e^2}{2\tau_S \tau_M} \right]^{1/2} = \left[ \frac{e^2}{4\tau_{STL} \tau_M} \right]^{1/2} = \frac{e}{2}\Gamma_0 \simeq 1.36\Gamma_0 \tag{63}$$

which is only slightly larger than the value found in the limit of $\tau_S \ll \tau_{STL}$. Thus, while operating at $\tau_S = 2\tau_{STL}$ drastically reduced the value of the MMFS due to the VM, it does not correspond to any significant increase in the value of the MMFS due to the SE.